\documentclass{aa}

\usepackage[utf8]{inputenc}
\usepackage[T1]{fontenc}

\usepackage[varg]{txfonts}

\usepackage{hyperref}
\usepackage{graphicx}   
\usepackage{amsmath}    
\usepackage{amssymb}    
\usepackage{threeparttable}
\usepackage{booktabs}
\usepackage{verbatim}
\usepackage[all]{hypcap}
\usepackage{fixmath}
\usepackage[usenames, dvipsnames]{color}

\begin{document}

\title{KiDS+GAMA: The weak lensing calibrated stellar-to-halo mass relation of central and satellite galaxies}

\author{Andrej Dvornik\inst{1,2}
\and Henk Hoekstra\inst{2}
\and Konrad Kuijken\inst{2} 
\and Angus H. Wright\inst{1}
\and Marika Asgari\inst{3}
\and Maciej Bilicki\inst{4}
\and Thomas Erben\inst{5}
\and Benjamin Giblin\inst{3}
\and Alister W. Graham\inst{6}
\and Catherine Heymans\inst{1,3}
\and Hendrik Hildebrandt\inst{1}
\and Andrew M. Hopkins\inst{7}
\and Arun Kannawadi\inst{2,8}
\and Chieh-An Lin\inst{3}
\and Edward N. Taylor\inst{6}
\and Tilman Tröster\inst{3}
}
\institute{Ruhr-University Bochum, Astronomical Institute, German Centre for Cosmological Lensing, Universitätsstr. 150, 44801 Bochum, Germany
\and Leiden Observatory, Leiden University, P.O. Box 9513, 2300 RA Leiden, the Netherlands  
\and Institute for Astronomy, University of Edinburgh, Royal Observatory, Blackford Hill, Edinburgh EH9 3HJ, UK
\and Center for Theoretical Physics, Polish Academy of Sciences, al. Lotników 32/46, 02-668 Warsaw, Poland
\and Argelander-Institut für Astronomie, Auf dem Hügel 71, 53121 Bonn, Germany
\and Centre for Astrophysics and Supercomputing, Swinburne University of Technology, Hawthorn 3122, Australia
\and Australian Astronomical Optics, Macquarie University, 105 Delhi Rd, North Ryde, NSW 2113, Australia
\and Department of Astrophysical Sciences, Princeton University, 4 Ivy Lane, Princeton NJ 08544, USA\\
\email{dvornik@astro.rub.de}}

\date{Received ???; Accepted ???}

\keywords{gravitational lensing: weak -- methods: statistical -- surveys -- galaxies: haloes -- dark matter -- large-scale structure of Universe.}

\titlerunning{KiDS+GAMA: Stellar-to-halo mass relation of central and satellite galaxies}
\authorrunning{A. Dvornik et al.}

\abstract{We simultaneously present constraints on the stellar-to-halo mass relation for central and satellite galaxies through a weak lensing analysis of spectroscopically classified galaxies. Using overlapping data from the fourth data release of the Kilo-Degree Survey (KiDS), and the Galaxy And Mass Assembly survey (GAMA), we find that satellite galaxies are hosted by halo masses that are $0.53 \pm 0.39$ dex (68\% confidence, $3\sigma$ detection) smaller than those of central galaxies of the same stellar mass (for a stellar mass of $\log(M_{\star}/M_{\odot}) = 10.6$). This is consistent with galaxy formation models, whereby infalling satellite galaxies are preferentially stripped of their dark matter. We find consistent results with similar uncertainties when comparing constraints from a standard azimuthally averaged galaxy-galaxy lensing analysis and a two-dimensional likelihood analysis of the full shear field. As the latter approach is somewhat biased due to the lens incompleteness and as it does not provide any improvement to the precision when applied to actual data, we conclude that stacked tangential shear measurements are best-suited for studies of the galaxy-halo connection.}

\maketitle

\section{Introduction}
\label{sec:intro}

According to the hierarchical galaxy formation model, galaxy groups and clusters form by the accretion of isolated galaxies and groups. This type of assembly process tidally strips mass from the infalling satellite galaxies and haloes. Because the dark matter is dissipationless (to a good approximation), it is more easily stripped from the subhalo than the baryons, which dissipate some of their energy and sink to the centre of their potential more efficiently than the dark matter, well before forming stars \citep{White1978}. Because the dark matter is not that centrally concentrated, it is thus more susceptible to tidal stripping than the baryons, even after a galaxy forms its stars, which are, to the first order, dissipationless as well. This model thus predicts that the satellite galaxies will be preferentially stripped of their dark matter and the effect can be observed as higher stellar mass to halo mass ratios of satellite galaxies compared to their central counterparts of a similar stellar mass. While some stars may be lost, relatively more dark matter will be stripped and the result is a higher stellar-to-halo mass (SMHM) ratio for satellites in dense environments, compared to centrals and/or less dense environments. Previous simulation studies \citep[see for example][]{Bower2006} show that the SMHM relation of satellite galaxies is significantly different from the SMHM relation of central galaxies. Further evidence for different SMHM relations for central and satellite galaxies comes from abundance matching methods, which show that the satellites typically have more stellar mass than the central galaxies for a given halo mass \citep{Reddick2013}, and they have similar or in some cases even larger stellar masses than at the infall time \citep[][and the references therein]{Reddick2013, Wechsler2018}, while their halo masses do not.
 
While the SMHM relation of central galaxies has been successfully measured by many studies \citep[for instance by][]{Hoekstra2005, Mandelbaum2006, More2011, VanUitert2011, Leauthaud2012}, this is not the case for satellite galaxies whose SMHM relation remains essentially unconstrained \citep{Sifon2018}. Recently, several weak gravitational lensing studies using galaxy groups and clusters have been undertaken \citep[such as the ones by][]{Limousin2007, Li2014a, Li2016, Sifon2015, Sifon2018}, all of them finding that the satellite galaxies' haloes are heavily truncated with respect to the central and field ones.

Weak gravitational lensing, through the lensing of background sources by a sample of galaxies, which is commonly called galaxy-galaxy lensing, directly measures the total mass of lensing galaxies, without assuming their dynamical state \citep{Bartelmann1999, Courteau2014}, and it is currently the only method available to measure the total mass of samples of galaxies directly. Measuring the lensing signal around satellite galaxies, however, can be particularly challenging for several reasons: their small contribution to the lensing signal by the host galaxy group, source blending at small separations, and sensitivity to field galaxy contamination \citep{Sifon2018}. As pointed out by \citet{Sifon2015}, the latter point is rather important as the field galaxies are not stripped and their contamination therefore complicates the interpretation of the lensing signal. 

These studies were based on tangentially averaging the shear, which washes out information for satellites to some extent. In \citet{Dvornik2019}, we revisit the two-dimensional galaxy-galaxy lensing method \citep{Schneider1997, Heymans2006}. This method, which tries to fit a two-dimensional shear field directly to the galaxy ellipticity measurements, was shown on simulated data to perform significantly better than the traditional one-dimensional analysis of stacked tangential shear profiles or the closely related excess surface density (ESD). One important advantage of the two-dimensional method lies in the fact that it exploits all of the information regarding the actual image configuration (the model predicts the shear for each individual background galaxy image) using the galaxies' exact positions, ellipticities, magnitudes, luminosities, stellar masses, group membership information, etc., rather than only using the ensemble properties of statistically equivalent samples \citep{Schneider1997}. Moreover, the clustering of the lenses is naturally taken into account, although it is more difficult to account for the expected diversity in density profiles \citep{Hoekstra2013book}. 

This method went out of fashion due to the unavailability of galaxy grouping information that would accurately classify galaxies as centrals and satellites \citep{Hoekstra2013book}, that is, the same information needed to robustly study the stellar mass to halo mass relation of satellite galaxies \citep{Sifon2015}. Treating the galaxies as centrals and satellites in a statistical way when considering the stacked signal could be naturally accounted for with the halo model \citep{Seljak2000, Peacock2000, Cooray2002}, thus overcoming the observational shortcomings. In recent years, this type of galaxy grouping information has become available thanks to the power of overlapping wide-field photometric surveys with highly complete spectroscopic surveys which allow one to treat central and satellite galaxies deterministically. The Kilo-Degree Survey \citep[KiDS,][]{Kuijken2015, DeJong2015} in combination with the overlapping Galaxy And Mass Assembly survey \citep[hereafter GAMA,][]{Driver2011, Robotham2011} provide an optimal data set for this type of analysis.

In this paper we present a two-dimensional galaxy-galaxy lensing measurement of the stellar-to-halo mass relation for central and satellite galaxies by combining a sample of spectrocopically confirmed galaxy groups from the GAMA survey and background galaxies from the fourth data release of KiDS \citep{Kuijken2019}. We use these measurements to constrain the stellar-to-halo mass relation, comparing the standard one-dimensional stacked tangential shear profiles with the two-dimensional galaxy-galaxy lensing method from \citet{Dvornik2019}.

The outline of this paper is as follows. In Sec. \ref{sec:data} we present the lens and source samples used in this analysis. In Sec. \ref{sec:lens_model} we present the specific lens model used in the paper and in Sec. \ref{sec:mla} we describe the two-dimensional galaxy-galaxy lensing formalism. The parameter inference procedure is presented in Sec. \ref{sec:fit}. We show the results in Sec. \ref{sec:results}, compare our results with the literature in Sec. \ref{sec:comparison}, and conclude with Sec. \ref{sec:conclusions}. Throughout the paper we use the following cosmological parameters, which enter into the calculation of the distances and other relevant properties \citep{PlanckCollaboration2014}: $\Omega_{\text{m}} = 0.307$, $\Omega_{\Lambda} = 0.693$, $\sigma_{8} = 0.8288$, $n_{\text{s}} = 0.9611$, $\Omega_{\text{b}} = 0.04825$, and $h = 0.6777$. The halo masses are defined as $M = 4\pi r_{\Delta}^3 \Delta \; \overline{\rho}_{\text{m}} / 3 $, the mass enclosed by the radius $r_\Delta$ within which the mean density of the halo is $\Delta$ times the mean density of the Universe $ \overline{\rho}_{\mathrm{m}}$, with $\Delta = 200$. All of the measurements presented in the paper are in co-moving units.

\section{Data and sample selection}
\label{sec:data}

\begin{figure}
        \includegraphics[width=\columnwidth]{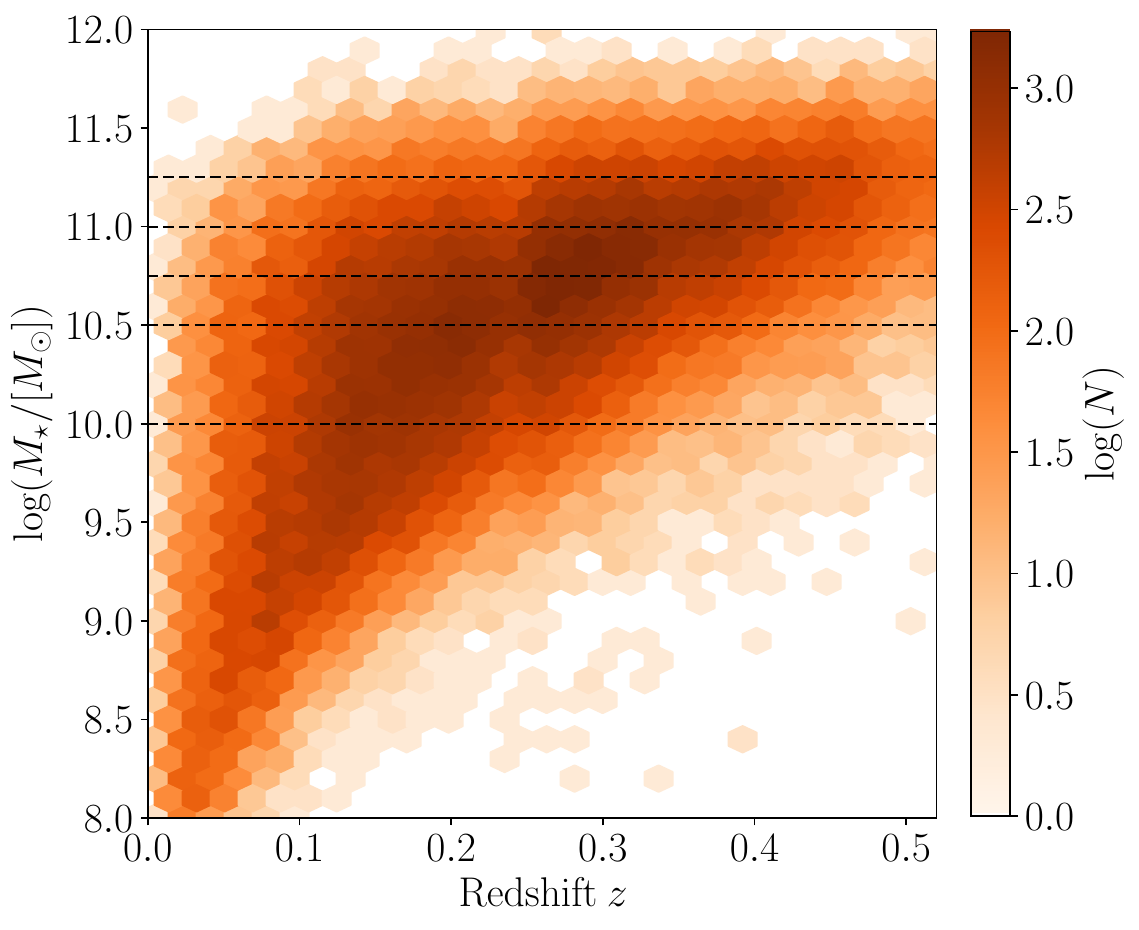}
        \caption{Stellar mass versus redshift of galaxies in the equatorial regions of the GAMA survey that overlap with KiDS. The full sample is shown with the hexagonal density plot and the dashed lines show the cuts for the stellar mass bins used in our analysis.}
        \label{fig:sample}
\end{figure}

\begin{figure}
        \includegraphics[width=\columnwidth]{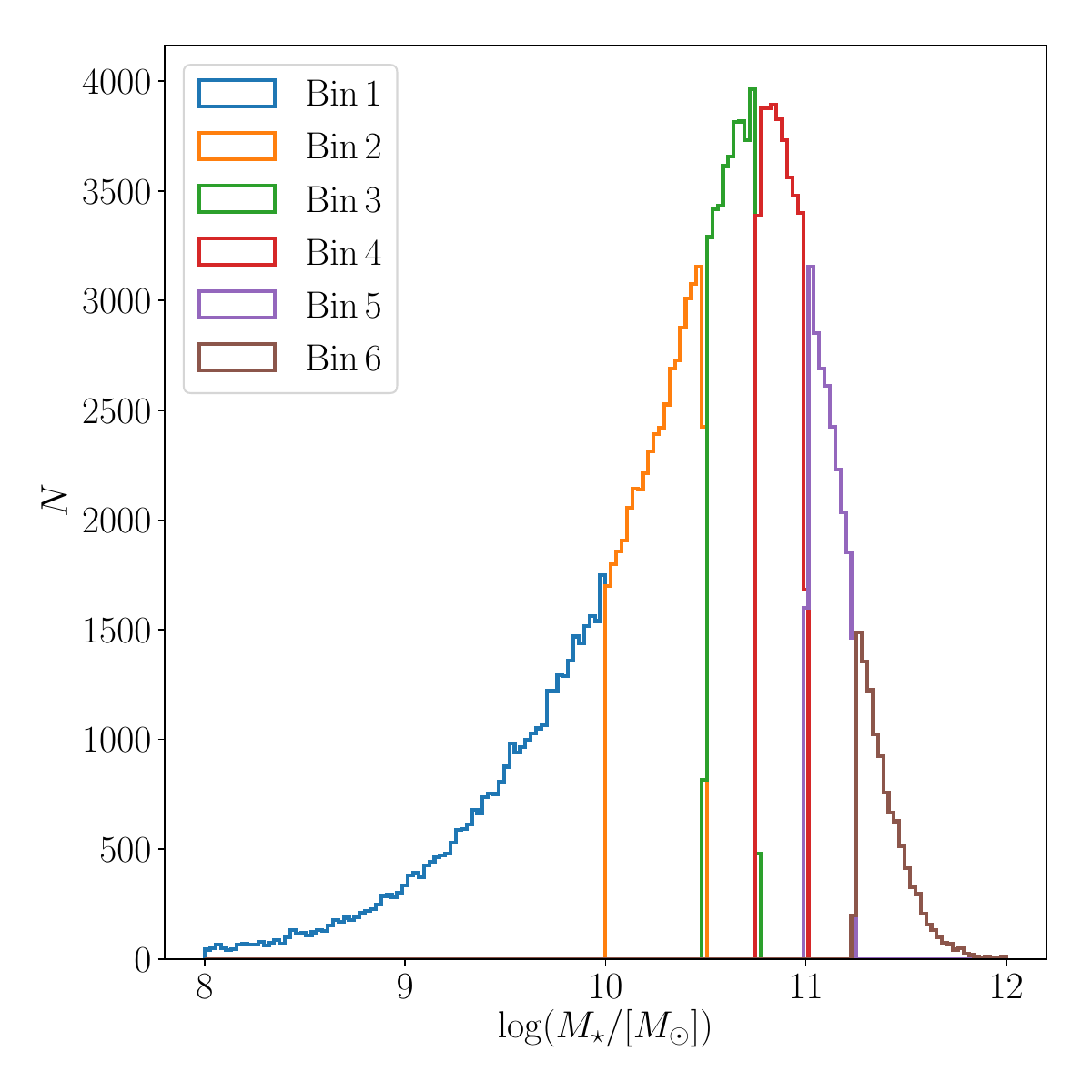}
        \caption{Stellar mass distributions in our six bins used for one-dimensional stacked tangential shear measurements. The exact bin values are presented in Table \ref{tab:sample}.}
        \label{fig:dists}
\end{figure}

The foreground galaxies used in this lensing analysis are taken from GAMA, a spectroscopic survey carried out on the Anglo-Australian Telescope with the AAOmega spectrograph. Specifically, we use the information of GAMA galaxies from three equatorial regions, G9, G12, and G15 from GAMA II \citep{Liske2015}. We do not use the G02 and G23 regions, as G02 does not overlap with KiDS and G23 uses an inconsistent target selection. These equatorial regions encompass \mbox{\textasciitilde{} 180 deg$^2$}, contain $180\,960$ galaxies (with $nQ \geq 3$, where the $nQ$ is an indicator of redshift quality), and are highly complete down to a Petrosian $r$-band magnitude of $r = 19.8$. We make use of the GAMA galaxy group catalogue by \citet{Robotham2011}, which provides information on the galaxy's group membership which is used to separate them into central and satellite galaxies. The GAMA galaxy group catalogue was constructed using a three-dimensional Friends-of-Friends (FoF) algorithm, linking galaxies in projected and line-of-sight separation. We use version 10 of the group catalogue (G3Cv10), which contains $26\,194$ groups with at least two members. All of the galaxies that are not grouped in any of the $26\,194$ groups are considered as centrals, which is shown to be a correct assumption in \citet{Brouwer2016a}. The GAMA survey is 98\% complete down to the observed magnitude limit. The group catalogue purity reaches 90\% for high multiplicity groups \citep{Robotham2011}, out of which 70 -- 75\% of centrals are correctly identified \citep{Robotham2011, Sifon2015}. We consider all galaxies whose stellar mass is between $10^{8} M_{\odot}$ and $10^{12} M_{\odot}$. Stellar masses are taken from version 20 of the \texttt{LAMBDAR} stellar mass catalogue, described in \citet{Wright2017}. The final selection of galaxies can be seen in Fig. \ref{fig:sample} and Fig. \ref{fig:dists}, and all of the relevant properties we need in our analysis are presented in Table \ref{tab:sample}. The stellar mass binning is only used for the one-dimensional galaxy-galaxy lensing case in order to obtain stacks of tangential shear signal, and it is chosen in such way that we have a similar signal-to-noise ratio in each stellar mass bin. In the two-dimensional case, we directly use the relevant individual galaxy quantities in the model.

\begin{table}
        \caption{Overview of the number of lens galaxies, median stellar masses of the galaxies, and median redshifts in each selected mass bin used for our one-dimensional stacked tangential shear analysis. Stellar masses are given in units of $\left[\log(M_{\star} / [M_{\odot}])\right]$.}
        \label{tab:sample}
        \begin{small}
        \centering
        \begin{tabular}{lrrrrrr} 
                \toprule
                Bin & \multicolumn{1}{c}{Range} & \multicolumn{1}{c}{$N_{\mathrm{tot}}$} &  \multicolumn{1}{c}{$N_{\mathrm{cen}}$} & \multicolumn{1}{c}{$N_{\mathrm{sat}}$} & \multicolumn{1}{c}{$M_{\star, \mathrm{med}}$} & \multicolumn{1}{c}{$z_{\mathrm{med}}$} \\ 
                \midrule
                1 & (8.0,10.0] & 39\,012 & 25\,908 & 13\,104 & 9.61 & 0.122 \\
                2 & (10.0,10.5] & 45\,416 & 28\,725 & 16\,691 & 10.29 & 0.193 \\
                3 & (10.5,10.75] & 34\,027 & 20\,819 & 13\,208 & 10.63 & 0.245 \\
                4 & (10.75,11.0] & 34\,714 & 20\,332 & 14\,382 & 10.87 & 0.285 \\
                5 & (11.0,11.25] & 22\,908 & 12\,594 & 10\,314 & 11.10 & 0.324 \\
                6 & (11.25,12.0] & 10\,705 & 5\,468 & 5\,237 & 11.36 & 0.380 \\
                \bottomrule
        \end{tabular}
        \end{small} 
\end{table}

We use imaging data from the $180$ deg$^2$ of the fourth   KiDS data release \citep{Kuijken2019} that overlaps with the three equatorial patches of the GAMA survey to obtain shape measurements of background galaxies. KiDS is a four-band imaging survey conducted with the OmegaCAM CCD mosaic camera mounted at the Cassegrain focus of the VLT Survey Telescope (VST); the camera and telescope combination provide us with a fairly uniform point spread function across the field-of-view. The companion VISTA-VIKING \citep{Edge2013} survey has provided complementary imaging in near-infrared bands ($Z$, $Y$, $J$, $H$, $K_{\mathrm{s}}$), resulting in a deep, wide, nine-band imaging dataset \citep{Wright2018}.

We use shape measurements based on the $r$-band images, which have an average seeing of $0.66$ arcsec. The image reduction, photometric redshift calibration, and shape measurement analysis is described in detail in \citet{Hildebrandt2018} and \citet{Kuijken2019}. We measure galaxy shapes using \emph{lens}fit \citep{Miller2013}, which has been calibrated using image simulations described in \citet{Kannawadi2018}. This provides galaxy ellipticities (${\epsilon_{1}}$, ${\epsilon_{2}}$) with respect to an equatorial coordinate system, and an optimal weight. 

\section{Lens model}
\label{sec:lens_model}

The most widely assumed density profile for dark matter haloes is the Navarro–Frenk–White (NFW) profile \citep{Navarro1995}.
Using simple scaling relations, this profile can be matched to simulated dark matter haloes over a wide range of masses and was found to be consistent with observations \citep{Navarro1995}. The NFW profile is defined as:
\begin{equation}
\rho_\mathrm{NFW}(r) = \frac{\delta_\mathrm{c}\, \overline{\rho}_{\mathrm{m}}}{(r/r_\mathrm{s})\, (1+r/r_\mathrm{s})^2},
\end{equation}
where the free parameters $\delta_\mathrm{c}$ and $r_\mathrm{s}$ are called the overdensity and the scale radius, respectively, $r$ is the radius, and $\overline{\rho}_{\mathrm{m}}$ is the mean density of the Universe, where $ \overline{\rho}_{\mathrm{m}} = \Omega_{\mathrm{m}} \rho_\mathrm{c}$ and $\rho_\mathrm{c}$ is the critical density of the Universe, defined by
\begin{equation}
\rho_\mathrm{c} \equiv \frac{3H^{2}_{0}}{8 \pi G},
\end{equation}
where $H_{0}$ is the present-day Hubble parameter.

Some thought is warranted when choosing how to model stripped satellites galaxies. In numerical simulations, the satellite galaxies are heavily stripped by their host halo, but the effect of stripping on their density profile is not that severe. Even though tidal stripping removes mass from the outskirts of the halo, tidal heating causes the subhalo to expand, and the resulting density profile is similar in shape to that of a central galaxy which has not been subject to tidal stripping \citep{Hayashi2003}. Similarly, \citet{Mira2011} found that the NFW profile is a better fit than truncated profiles for subhaloes in the Millennium Simulation \citep{Springel2005}, and that the reduction in mass produced by tidal stripping is simply reflected as a change in the NFW concentration of subhaloes. Following \citet{Sifon2018}, we have decided to model both centrals and satellites using the NFW profile, but while allowing the concentrations to differ.

The NFW profile in its usual parametrisation has two free parameters for each halo, halo mass $M_{\mathrm{h}}$, and concentration $c$, and using those is the conventional way of modelling halo profiles. However, having two free parameters for each halo is computationally very expensive. Instead, we would like to describe these parameters through relations that depend on halo properties and then fit to a few free parameters in these global relations instead of hundreds or thousands of free, halo-specific parameters. To do so, we adopt the halo mass -- concentration relation of \citet{Duffy2011}, with the free concentration normalisation $f_{\mathrm{c}}$: 
\begin{equation}
\label{eq:con_duffy}
c(M_{\mathrm{h}}, z) = f_{\mathrm{c}} \,10.14\;  \left[\frac{M_{\mathrm{h}}}{(2\times 10^{12} M_{\odot}/h)}\right]^{- 0.081}\ (1+z)^{-1.01} \,.\end{equation}
We describe the stellar mass to halo mass relation as an exponential function:
\begin{equation}
\label{eq:smhm}
M_{\mathrm{h}} / M_{\odot}  = \left( \frac{\alpha - \log(M_{\star}/M_{\odot})}{e^{\gamma}}\right)^{\beta}\,, 
\end{equation}
where\footnote{We note that $\alpha$ is set empirically as the stellar-to-halo mass function diverges at that value, thus we set it at the value that is higher than the largest stellar mass in our sample.} $\alpha = 12.0$, and $\beta$ and $\gamma$ are the free parameters we will be fitting. We note that the functional form presented here stems from \citet{Matthee2016}, but we have redefined some of the quantities\footnote{Our parameter $\beta$ is the same quantity as $1/\beta\log(e))$ in \citet{Matthee2016}, from the equation in their footnote 4.}. We use separate relations for the central and satellite galaxies, thus we have two sets of $\beta$ and $\gamma$ parameters since we want to constrain the SMHM relation for those populations separately. The choice of this type of parametrisation for the SMHM relation is motivated by reducing the number of free parameters required for the fit, while to first order maintaining the shape of the relation that is similar to the more widely adopted double power law parametrisations \citep[as presented by][]{Leauthaud2012, Moster2013, vanUitert2016}.

The gravitational shear and convergence profiles are calculated using equations 14 to 16 from \citet{Wright1999}, from which the predicted ellipticities for all of the lenses are calculated according to the weak lensing relations presented in \citet{Schneider2003}. We first calculate the reduced shear for our NFW profiles:
\begin{equation}
\label{eq:red_shear}
g(x_{i}, z_{\mathrm{s}}) = \frac{\gamma(x_{i}, z_{\mathrm{s}})}{1-\kappa(x_{i}, z_{\mathrm{s}})}\,, 
\end{equation}
from which the ellipticities are calculated according to the following equation:
\begin{equation}
\epsilon =
        \begin{cases}
            g &\quad \textrm{if}\ \vert g \vert \leq 1 \\
            1/g^{*} &\quad \textrm{if}\ \vert g \vert > 1
        \end{cases}\,, 
\end{equation}
where we have assumed that the intrinsic ellipticities of the sources average to 0 due to their random nature. Intrinsic alignments are not thought to contribute significantly to the signal at the current signal-to-noise ratio \citep{Blazek2012}.

We compute the effective critical surface mass density that we need in our lens model for each lens using the spectroscopic redshift of the lens $z_{\mathrm{l}}$ and the full normalised redshift probability density of the sources, $n(z_{\mathrm{s}})$, calculated using the direct calibration method presented in \citet{Hildebrandt2016, Hildebrandt2018}. The effective inverse critical surface density\footnote{We refer the reader to \citet{Dvornik2018}, Appendix C for a full
discussion on the different definitions of $\Sigma_{\mathrm{cr}}$ that have been adopted in the literature.} can be written as:
\begin{equation}
\label{eq:crit_effective}
\Sigma_{\mathrm{cr, ls}}^{-1}=\frac{4\pi G}{c^2} (1+ z_{\rm l})^{2} D(z_{\mathrm{l}}) \int_{z_{\mathrm{l}}}^{\infty} \frac{D(z_{\mathrm{l}},z_{\mathrm{s}})}{D(z_{\mathrm{s}})}n(z_{\mathrm{s}})\, \mathrm{d}z_{\mathrm{s}} \, ,
\end{equation}
where $D(z_{\rm l})$ is the angular diameter distance to the lens, $D(z_{\rm l}, z_{\rm s})$ is the angular diameter distance between the lens and the source, and $D(z_{\rm s})$ is the angular diameter distance to the source.

The galaxy source sample is specific to each lens redshift with a minimum photometric redshift $z_{s} = z_{\mathrm{l}} +\delta_{z}$, with $\delta_{z} = 0.2$, where $\delta_{z}$ is an offset to mitigate the effects of contamination from the group galaxies \citep[for details, see also the methods section and Appendix of][]{Dvornik2017a}. We determine the source redshift distribution $n(z_{\mathrm{s}})$ for each sample by applying the sample photometric redshift selection to a spectroscopic catalogue that has been weighted to reproduce the correct galaxy colour-distributions in KiDS \citep[for details see][]{Hildebrandt2018}. The accuracy of this method, which is determined through mock data analysis, is sufficient for our study \citep{Wright2018}. We correct the measured ellipticities for the multiplicative shear bias per source galaxy per redshift bin as defined in \citet{Hildebrandt2018} with the (small) correction values estimated from image simulations \citep{Kannawadi2018}.

\section{Galaxy-galaxy lensing formalism}
\label{sec:mla}

In this study of satellite galaxy-galaxy lensing, we use the two-dimensional galaxy-galaxy lensing formalism as presented in \citet{Dvornik2019}; we follow their model and adapted it to KiDS+GAMA by taking into account the survey's specific requirements. Generally, for both the one-dimensional and two-dimensional cases, the likelihood of a model with a set of parameters $\mathbold{\theta}$ given data $\mathbf{d}$ can be parametrised in the following form:
\begin{equation}
\mathcal{L}(\mathbf{d}\, \vert\, \mathbold{\theta}) =  \frac{1}{\sqrt{\left(2 \pi \right)^{n} \vert \mathbf{C} \vert}} \exp \left[-\frac{1}{2} \left(\mathbf{m} (\mathbold{\theta}) - \mathbf{d} \right)^{T}{\mathbf{C}}^{-1} \left(\mathbf{m} (\mathbold{\theta}) - \mathbf{d}\right) \right],
\end{equation}
where $\mathbf{m}(\mathbold{\theta})$ is the value of $\mathbf{d}$ predicted by the model with parameters $\mathbold{\theta}$. We assume the measured data points $\mathbf{d} = [d_{1}, \dots, d_{n}]$ are drawn from a normal distribution with a mean equal to the true values of the data, and $n$ is the dimensionality of the data. In principle, the likelihood does not need to be Gaussian, but in practice it is a very good approximation due to ellipticity distribution being nearly Gaussian as well. The likelihood function accounts for correlated data points through the covariance matrix $\mathbf{C}$ of shape $n\times n$. The covariance matrix $\mathbf{C}$ generally consists of two parts, the first arising from shape noise and the second from the presence of a cosmic structure between the observer and the source \citep{Hoekstra2003a}:
\begin{equation}
\mathbf{C} = \mathbf{C}^{\mathrm{shape}} + \mathbf{C}^{\mathrm{LSS}} \,.
\end{equation} 
In the case when one wants to fit one-dimensional tangential shear profiles, which are stacked over a sample of lenses, the likelihood function can be written as:
\begin{align}\label{eq:tangential}
\mathcal{L}(g_{\mathrm{t}}^{\mathrm{obs}}  \, \vert \, M_{\mathrm{h}}, M_{\star}, c) & \\ \nonumber
= \prod_{i=1}^{n} \frac{1}{\sigma_{g_{\mathrm{t}}, i} \sqrt{2 \pi}} &\exp \left[-\frac{1}{2} \left(\frac{g_{\mathrm{t}, i} (M_{\mathrm{h}}, R_{i}, z)  - g_{\mathrm{t}, i}^{\mathrm{obs}}}{\sigma_{g_{\mathrm{t}}, i}}\right)^{2} \right],
\end{align}
where we have used $m_{i} = g_{\mathrm{t}, i} (M_{\mathrm{h}}, R_{i}, z)$ (see Eq. \ref{eq:red_shear}) as the model prediction given halo mass $M_{\mathrm{h}}$, radial bin $R_{i}$, and redshift of the lens $z$, and $d_{i} = g_{\mathrm{t}, i}^{\mathrm{obs}}$ as the tangentially averaged reduced shear of a sample of lenses measured from observations. The halo mass $M_{\mathrm{h}}$ and stellar mass $M_{\star}$ are connected through Eq. \ref{eq:smhm}, and the concentration $c$ is defined in Eq. \ref{eq:con_duffy}. Here we have also used the uncertainty of our measurement, given by the $\sigma_{g_{\mathrm{t}}, i}$ calculated from the intrinsic shape noise of sources in each radial bin. The product runs over all radial annuli $i$. Moreover we only account for the diagonal terms of the covariance matrix and only include the error due to the shape noise, that is to say
\begin{equation}
        \sqrt{|C|} = \prod_{i=1}^{n} \sigma_{i} \,.
\end{equation}
This is done to reduce the computational complexity of the problem, and it is justified due to the covariance matrix being shape noise dominated. The above equation describes the one-dimensional method; the two-dimensional method differs only in the following significant way:
\begin{align}\label{eq:max_like}
\mathcal{L}(\epsilon^{\mathrm{obs}} \, \vert \, M_{\mathrm{h}}, M_{\star}, c) & \\ \nonumber
= \prod_{i=1}^{n} \frac{1}{\sigma_{\epsilon, i} \sqrt{2 \pi}} &\exp \left[-\frac{1}{2} \left(\frac{g_{i} (M_{\mathrm{h}}, x_{i}, z) - \epsilon_{i}^{\mathrm{obs}}}{\sigma_{\epsilon, i}}\right)^{2} \right],
\end{align}
where $g_{i} (M_{\mathrm{h}}, x_{i}, z)$ are the model reduced shears evaluated at each source position $x_{i}$, $\epsilon_{i}^{\mathrm{obs}}$ is the observed elipticities of real galaxies, and $\sigma_{\epsilon, i}$ is the intrinsic shape noise of our galaxy sample per component, calculated from the \emph{lens}fit weights following the description by \citet{Heymans2012}. The same \emph{lens}fit weights are used to weight $\epsilon_{i}^{\mathrm{obs}}$ as well \citep{Heymans2012}. In practice, the two-dimensional fit to the ellipticities is carried out for each Cartesian component of ellipticity $\epsilon_{1}$ and $\epsilon_{2}$ with respect to the equatorial coordinate system. Here the product runs over all of the individual source galaxies $i$.

\section{Parameter inference procedure}
\label{sec:fit}

Due to the computational complexity of the analysis, we do not use the Markov chain Monte Carlo (MCMC) method for parameter inference for any of our methods. Our model parameter inference procedure and the fit to the data is performed using three steps for both the one-dimensional and two-dimensional cases. We first sample 100 points using the latin hypercube method \citep{McKay1979} in the six-dimensional model parameter space (see Table \ref{tab:results} for the ranges of all of the parameters). We picked these types of ranges for the parameters in order to sample the likelihood surface in the $5\sigma$ range that we found using an MCMC fit of the 1D model in preliminary tests. This minimises the need for having a larger number of points in the latin hypercube as well as reducing the number of interpolation points at the later step. For each of the 100 points, we calculate the likelihood value $\mathcal{L}(\mathbf{d}\, \vert\, \mathbold{\theta})$ according to equations \ref{eq:tangential} and \ref{eq:max_like}. We created $10\,000$ realisations of the latin hypercube and select one that maximises the euclidean distance between all of the points \citep[similarly to what it was done by][]{Heitmann2009}. We have also checked the influence of different realisations on the obtained results by evaluating the one-dimensional method on different realisations, and we always obtain the same resulting likelihoods.

\begin{figure}
        \includegraphics[width=\columnwidth]{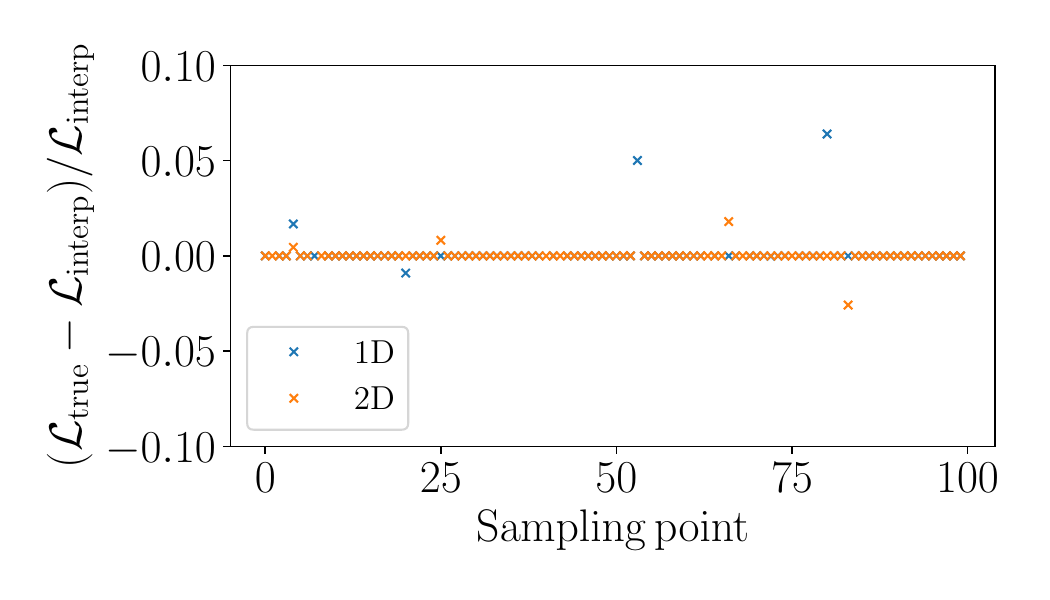}
        \caption{Fractional differences between the interpolated likelihood and true likelihood value at each sampled point in the latin hypercube. We do not show the cases where one of the interpolated points is exactly on one of the edges of the latin hypercube since the interpolator is unable to properly perform for those edge cases. The remaining high value outliers are the points in the latin hypercube that are close to the edges of the parameter space.}
        \label{fig:rbf}
\end{figure}

The second step requires the construction of an interpolator, for which we use the Gaussian process (GP) regression method with a multi-dimensional radial basis function (RBF) kernel\footnote{We compute GP interpolation using the \texttt{scikit-learn} package (\url{http://scikit-learn.org}).}. The interpolated value is the likelihood $\mathcal{L}(\mathbf{d}\, \vert\, \mathbold{\theta})$ as obtained from the latin-hypercube samples. Interpolation is performed with the same ranges as used in the construction of the latin hypercube using $20^6$ ($64$ million) equally spaced grid points. We test the accuracy of the interpolation by choosing one point as a test point, and we use the remaining 99 likelihoods to construct the interpolator. We then compare the prediction at the test point to the actual likelihood value. This is repeated for all likelihood values of our latin hypercube. We show the fractional differences between the interpolated likelihood and true likelihood value at each point in Fig. \ref{fig:rbf}. The majority of sampled points are accurate to better than 1\%. The use of the latin hypercube to construct the initial grid to sample the likelihood surface does not increase the uncertainty in recovering the true values due to the properties of latin hypercube sampling compared to the usual grid search minimisation \citep{McKay1979}.  

In the third step, we calculate the marginalised distributions from the interpolated points. First we normalise the probability grid $P_{\mathrm{6D}}$, such that:
\begin{equation}
\sum_{\mathbold{\theta}} P_{\mathrm{6D}} (\mathbf{d}\, \vert\, \mathbold{\theta}) = \sum_{\mathbold{\theta}} \mathcal{L}(\mathbf{d}\, \vert\, \mathbold{\theta}) =  1 \,,
\end{equation}
which also sets the normalisation term in Bayes' theorem, so we can, from the resulting values, calculate the one-dimensional and two-dimensional marginalised distributions of all of the parameters.

\section{Results}
\label{sec:results}

\begin{figure}
        \includegraphics[width=\columnwidth]{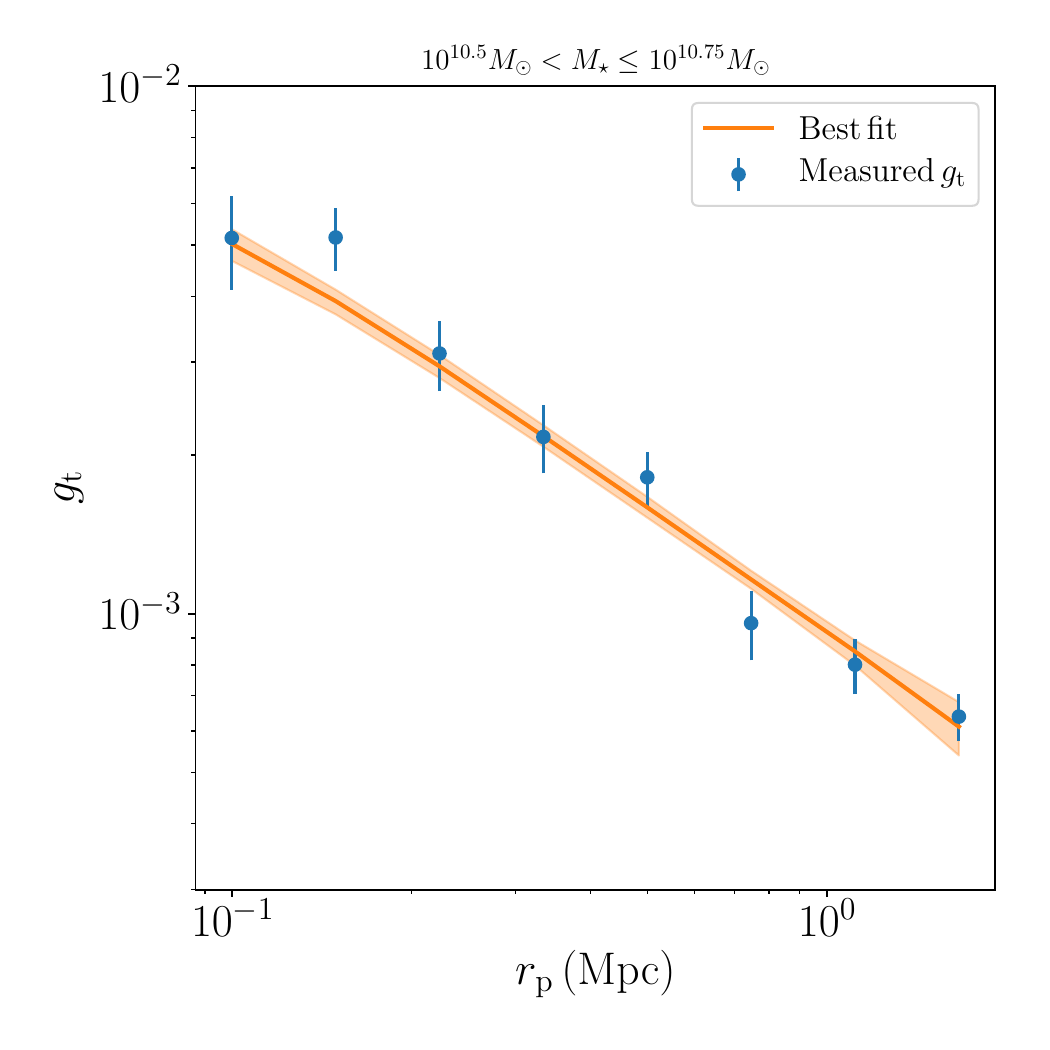}
        \caption{Stacked tangential shear profile for the GAMA lenses (blue points) in the $10^{10.5}$ to $10^{10.75} M_{\odot}$ stellar mass bin, compared to the best fitting lensing model for the 1D method, with contributions from both centrals and satellites. The orange band encloses the 68\% credible interval.}
        \label{fig:result2}
\end{figure}

\begin{figure*}
        \centering
        \includegraphics[width=\textwidth]{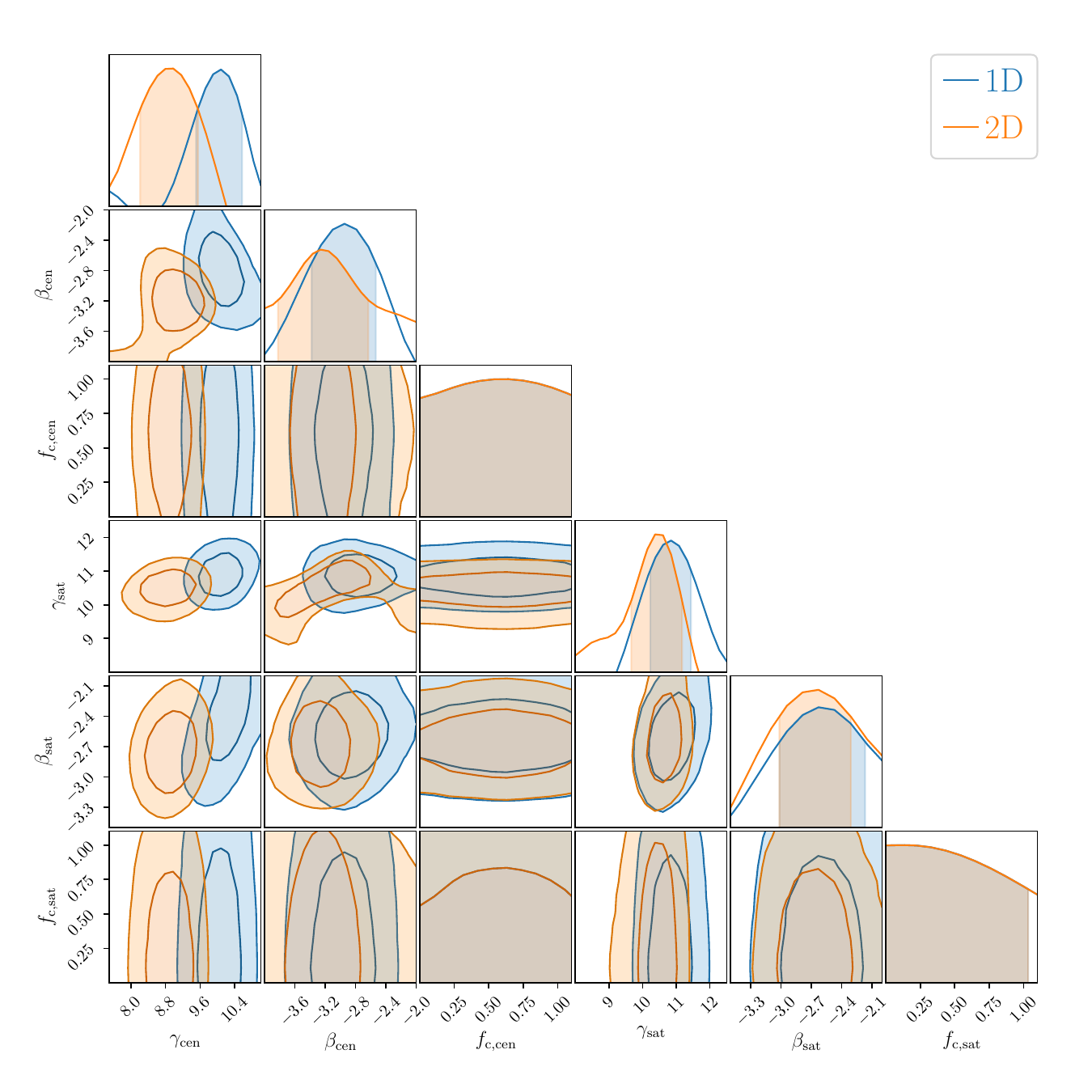}
        \caption{Full posterior distributions of the model parameters $\gamma_{\mathrm{cen}}$, $\beta_{\mathrm{cen}}$, $f_{\mathrm{c, cen}}$, $\gamma_{\mathrm{sat}}$, $\beta_{\mathrm{sat}}$, and $f_{\mathrm{c, sat}}$, for both the one-dimensional stacked tangential shear measurements (in blue) as well as the two-dimensional galaxy-galaxy lensing method (in orange). The contours indicate the $1\sigma$ and $2\sigma$ credible regions. We note that $\gamma$ and $\beta$ are the free parameters of the SMHM relation (Eq. \ref{eq:smhm}) and $f_{\mathrm{c}}$ is the concentration-mass relation normalisation parameter (Eq. \ref{eq:con_duffy}).}
        \label{fig:corner}
\end{figure*}

We fit the lens model as described in Sect \ref{sec:lens_model} to the stacked tangential shear measurements in our six stellar mass bins (our 1D result) and to the full two-dimensional shear field (our 2D result). An example of a single stacked tangential shear profile for the GAMA lenses in the $10^{10.5}$ to $10^{10.75} M_{\odot}$ stellar mass bin is shown in Fig. \ref{fig:result2}, with the measurements and their respective $1\sigma$ errors\footnote{For our two-dimensional analysis, there is no corresponding visualisation for the data vector $d$, other than the noisy residual shear field, which is not shown because it is not pertinent.}. The measured lens model best-fit parameters (median of the marginalised posterior estimate), together with their 68\% credible intervals are presented in Table \ref{tab:results} for both the one-dimensional and two-dimensional analysis. The constrains from the two approaches can be compared in terms of the full posterior distributions shown in Fig. \ref{fig:corner}. Even though none of our parameters are constrained to within $5\sigma$ of the preliminary test, the prior ranges are still good, given that models outside of the prior range for $\gamma$ and $\beta$ parameters would be unphysical.

\begin{table}
    \centering
    \caption{Parameter space ranges and marginalised posterior estimates of the free parameters used in our lens model, for both the one-dimensional method and the two-dimensional method. We note that $\gamma$ and $\beta$ are the free parameters of the SMHM relation (Eq. \ref{eq:smhm}) and $f_{\mathrm{c}}$ is the concentration-mass relation normalisation parameter (Eq. \ref{eq:con_duffy}). Furthermore, $f_{\mathrm{c}}$ parameters recover the prior range, thus we do not provide their values.}
    \label{tab:results}
    \setlength\tabcolsep{4pt}
    \begin{tabular}{cccc}
        \toprule
                ~ & $\gamma_{\mathrm{cen}}$ & $\beta_{\mathrm{cen}}$ & $f_{\mathrm{c, cen}}$ \\  
               \midrule
                Parameter range & $[7.5, 11]$ & $[-5, -2]$ & $[0, 1.1]$ \\
                                \addlinespace
                1D results & $10.08^{+0.49}_{-0.57}$ & $-2.95^{+0.42}_{-0.43}$ & -- \\ 
                \addlinespace
                2D results& $8.97^{+0.58}_{-0.76}$ & $-3.26^{+0.63}_{-0.56}$ & -- \\ 
                \midrule
                ~ & $\gamma_{\mathrm{sat}}$ & $\beta_{\mathrm{sat}}$ & $f_{\mathrm{c, sat}}$ \\
                \midrule
                Parameter range & $[8, 12.5]$ & $[-5, -2]$ & $[0, 1.1]$ \\
                                \addlinespace
                                1D results & $10.84^{+0.60}_{-0.61}$ & $-2.63^{+0.46}_{-0.38}$ & -- \\ 
                \addlinespace
                2D results & $10.37^{+0.81}_{-0.70}$ & $-2.63^{+0.32}_{-0.39}$ & -- \\
                
                \bottomrule
    \end{tabular}
\end{table}

For both methods, we find that the SMHM relations are completely described with two parameters each (two for centrals and two for satellites): the normalisation $\gamma$ and slope $\beta$, for which the obtained one-dimensional values for centrals and satellites are presented in Table \ref{tab:results}. In the same table, we present the obtained values for centrals and satellites in the case when using the two-dimensional galaxy-galaxy lensing. These constraints are consistent with those from the one-dimensional analysis with the highest discrepancy in the $\gamma_{\mathrm{cen}}$ parameter, which differs by $1\sigma$. This shows that the two methods perform equally well statistically. 

The normalisations of the concentration-halo mass relation are essentially unconstrained in the adopted prior range. The other values of the SMHM relation are not correlated with these values, and the prior does not influence our results in a substantive way. The wide range is somewhat driven by the imperfect modelling for both the one-dimensional and the two-dimensional case, as the model does not properly account for the two-halo term. The prior ranges are comparable to the values found in hydrodynamical simulations \citep{Dvornik2019}. These values are also consistent with the observational findings that prefer lower normalisations than expected in simulations, such as in the studies of \citet{Viola2015}, \citet{Sifon2015}, and \citet{Dvornik2017a}. Since there are no strong covariances between $f_{\mathrm{c}}$ and the other parameters, any small systematic error in $f_{\mathrm{c}}$ probably does not propagate through to a bias in other parameters of the SMHM relation.

\begin{figure}
        \includegraphics[width=\columnwidth]{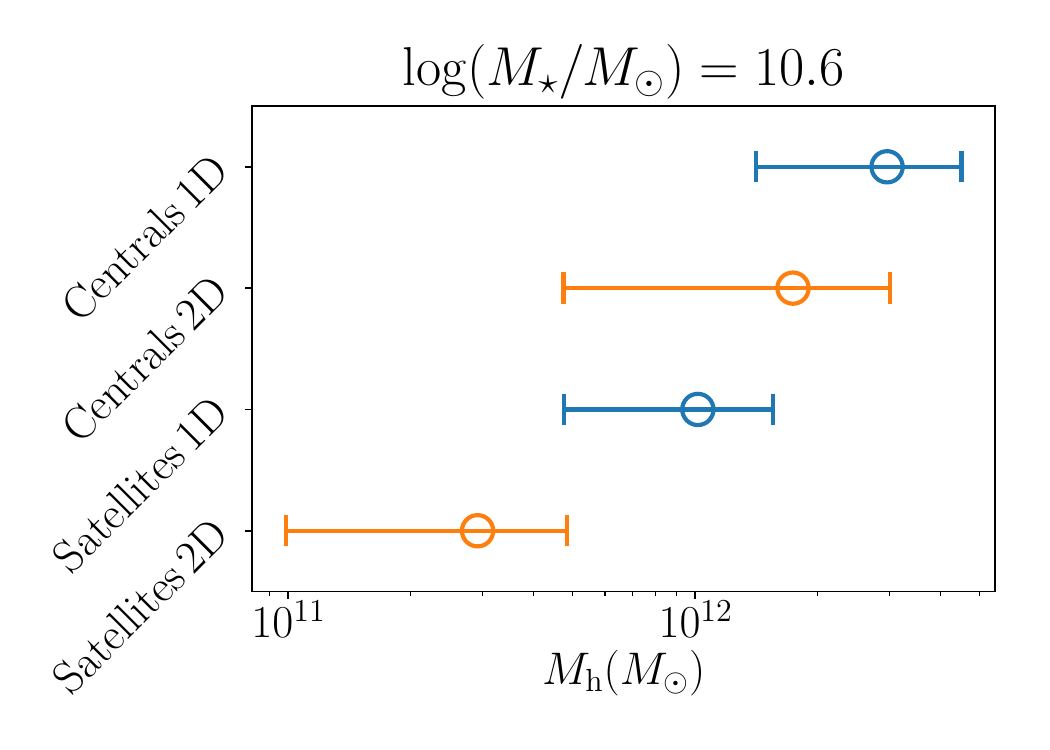}
        \caption{Halo masses for a galaxy with a stellar mass of $\log(M_{\star}/M_{\odot}) = 10.6$ for the one-dimensional and two-dimensional analyses for both central and satellite galaxies.}
        \label{fig:smhm1}
\end{figure}

We show a typical halo mass for a central and satellite galaxy with a stellar mass of $\log(M_{\star}/M_{\odot}) = 10.6$ in Fig. \ref{fig:smhm1}, obtained from propagating the best fit parameters through the SMHM relation. We find that the SMHM relations are different for the central and satellite galaxies, showing that the stripping of the dark matter does indeed take place; the SMHM relation of satellite galaxies is higher than the relation for the centrals, as is also seen in Fig. \ref{fig:smhm1}. We find that satellite galaxies are hosted by halo masses that are systematically $0.53  \pm 0.39$ dex (for 1D) and $0.23 \pm 0.18$ dex (for 2D) smaller than those of central galaxies at this stellar mass. The uncertainty of the inferred SMHM relation is similar to the intrinsic scatter present in simulations, for instance by the EAGLE hydrodynamical simulation \citep[][see Fig. \ref{fig:smhm_cen_comp} and Fig. \ref{fig:smhm_sat_comp}]{Schaye2015, Matthee2016}. While we see the same qualitative conclusions between the one-dimensional and two-dimensional analysis, the quantitative halo masses inferred are inconsistent at the \textasciitilde{}$1\sigma$ level.

\subsection{Assessment of completeness bias in 2D galaxy-galaxy lensing}

As is shown in \citet{Dvornik2019}, the two-dimensional analysis relies on a complete sample of lenses. If lenses are missing from the model, the bias in halo mass can be as much as 20\%. Although we used all of the galaxies with redshifts from the GAMA data, at a given redshift, the magnitude limit implies a limit in stellar mass. Galaxies with that stellar mass but at higher redshifts are not included in the catalogue, but they do contribute to the lensing signal. The contribution to the lensing signal is small for very high redshift lenses; however, not including lenses near the magnitude limit of GAMA may bias our measurements. Moreover, the process of labelling galaxies into centrals and satellites is not perfect. The GAMA group catalogue is only pure up to 90\% for high multiplicity groups, and it is known to be contaminated by the misidentification of the central galaxy in a group to such an extent that only 70 -- 75\% of central galaxies are correctly identified \citep{Robotham2011}. Thus the true central galaxy would be included in the satellite sample, which can introduce a bias of about 15\% in the inferred masses \citep{Sifon2015}. This effect is even more pronounced in the GAMA group catalogue for pairs of galaxies, where both components are likely to be centrals and not a central galaxy and one satellite galaxy. What is more, the satellite stellar-to-halo mass relation at a high stellar mass is possibly driven by this misidentification of satellite galaxies, which should actually be classified as centrals, given the high halo masses measured; satellite galaxies with stellar masses up to $\log(M_{\star}/M_{\odot}) = 12$ should not be common. This is a likely consequence of the observed problem with the FoF algorithm used to identify galaxy groups in the GAMA survey, but it does not seem to substantially affect the results. The FoF algorithm separates groups into a number of smaller groups or smaller, aggregate, unrelated groups into one large group, which would then host more than one central galaxy with them being classified as a satellite \citep{Jakobs2018}. 

In order to assess the possible bias due to missing galaxies in a magnitude limited survey, such as GAMA, we use the MICE-GC N-body simulation from which the MICE collaboration constructed a lightcone spanning a full octant on the sky \citep{Fosalba2015a, Fosalba2015}. The MICE lightcone has a maximum redshift of 1.4. The haloes found in the simulation were populated using a hybrid halo occupation distribution (HOD) and halo abundance matching (HAM) prescription \citep{Carretero2015, Crocce2015, Hoffmann2015}. For our assessment, we use the MICECATv2.0 catalogue\footnote{The MICECATv2.0 catalogue is available through CosmoHub \url{https://cosmohub.pic.es}.}, from which we take the positions of galaxies within a 4 deg$^2$ cutout of the lightcone with a redshift limit of $z < 0.5$ and an SDSS r-band magnitude of $m_{\mathrm{r}} < 22$. We also select galaxies with a stellar mass between $10^{7}\, M_{\odot}/h$ and $10^{13}\, M_{\odot}/h$. This selection of galaxies results in a distribution of stellar masses and redshifts similar to that of GAMA, which also has a similar number density. On this sample of galaxies, we apply an additional magnitude cut of $m_{\mathrm{r}} < 19.8$, which is the magnitude limit of the GAMA survey \citep{Driver2011}. 

\begin{figure}
        \includegraphics[width=\columnwidth]{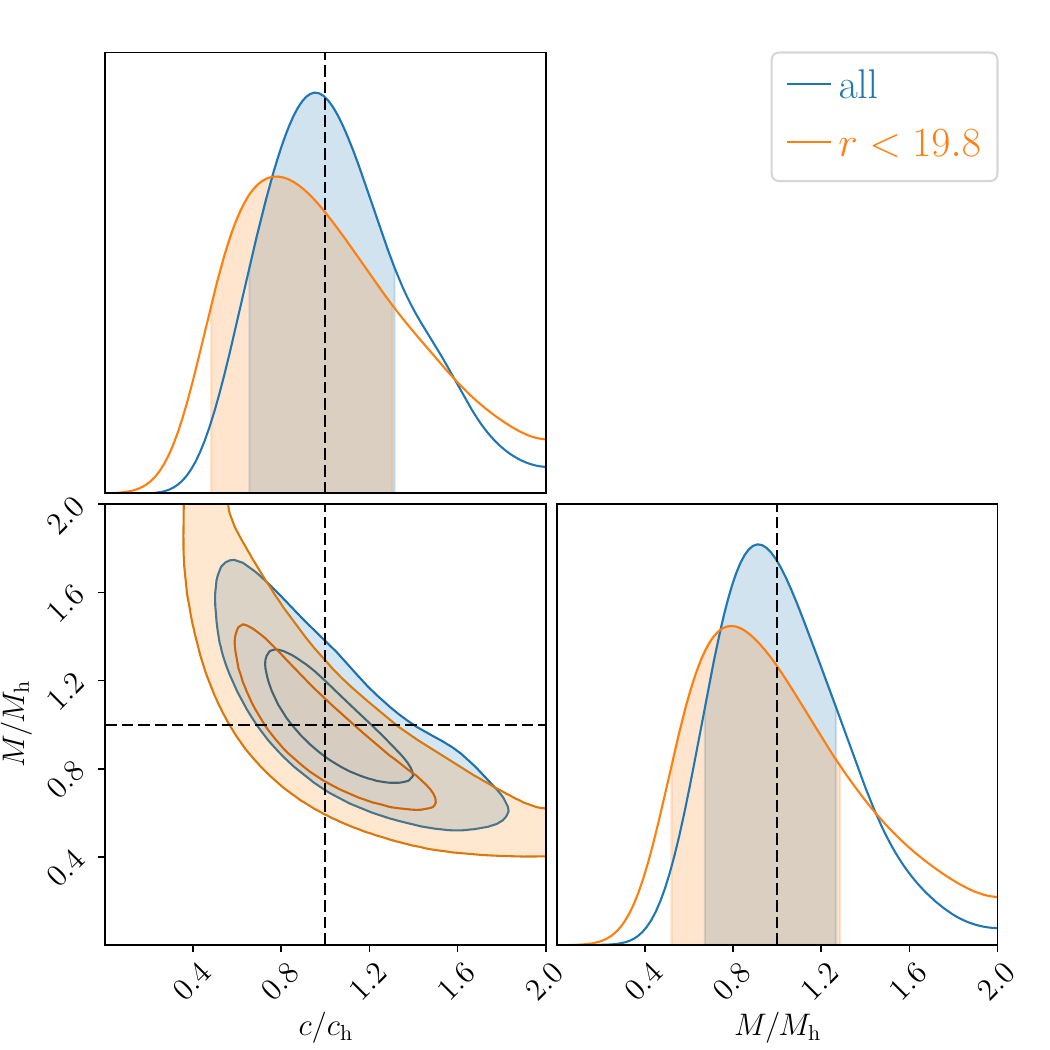}
        \caption{Comparison of the inferred parameters between the full sample of MICE galaxies (blue) and for galaxies with $m_{\mathrm{r}} < 19.8$ (orange). The model is able to accurately recover the input relation for the full sample of galaxies (dashed line), while the magnitude limited sample is biased at the 10\% level.}
        \label{fig:mice}
\end{figure}

We generate a noiseless mock shear field resembling a typical KiDS observation using the procedure shown in \citet{Dvornik2019}. We populate the mock shear field with haloes at the locations of galaxies from MICE mocks by assigning the stellar-to-halo mass relation from \citet{Matthee2016}, using the redshifts we have in the MICE mocks. We fit for the concentration and SMHM normalisations, using the two mock samples (the $m_{\mathrm{r}} < 22$ and the $m_{\mathrm{r}} < 19.8$ magnitude limited samples) as our input lenses for the fits. The parameter inference method is the same as described in Sec. \ref{sec:fit}. We show the comparison of the inferred parameters in Fig. \ref{fig:mice}, between the full sample of MICE galaxies (blue) and analysis for lens galaxies with $m_{\mathrm{r}} < 19.8$ (orange). The model is able to accurately recover the input relation for the full sample of galaxies, while that is no longer the case for a magnitude limited sample. Due to a smaller number of lenses, the uncertainty increases but also the retrieved concentrations and halo masses are biased towards lower values. The effect is present at the 10\% level, which is consistent with what we have found for a mock dataset of randomly placed lenses at a fixed redshift \citep{Dvornik2019}, but it is more representative of a real galaxy distribution and the observed effects due to the magnitude limit. The 10\% change seems to be consistent with a statistical fluctuation, given that the recovered values lie within the $1\sigma$ contours. We do need to point out that this test is performed on noiseless data, for which we can choose the size of the contours, and the change in values reflects the true bias. Given the bias in the two-dimensional analysis and the comparable statistical performance of the two methods, our one-dimensional constraints are our preferred result for this analysis. 

\section{Comparison with previous studies}
\label{sec:comparison}

\begin{figure*}
        \includegraphics[width=\textwidth]{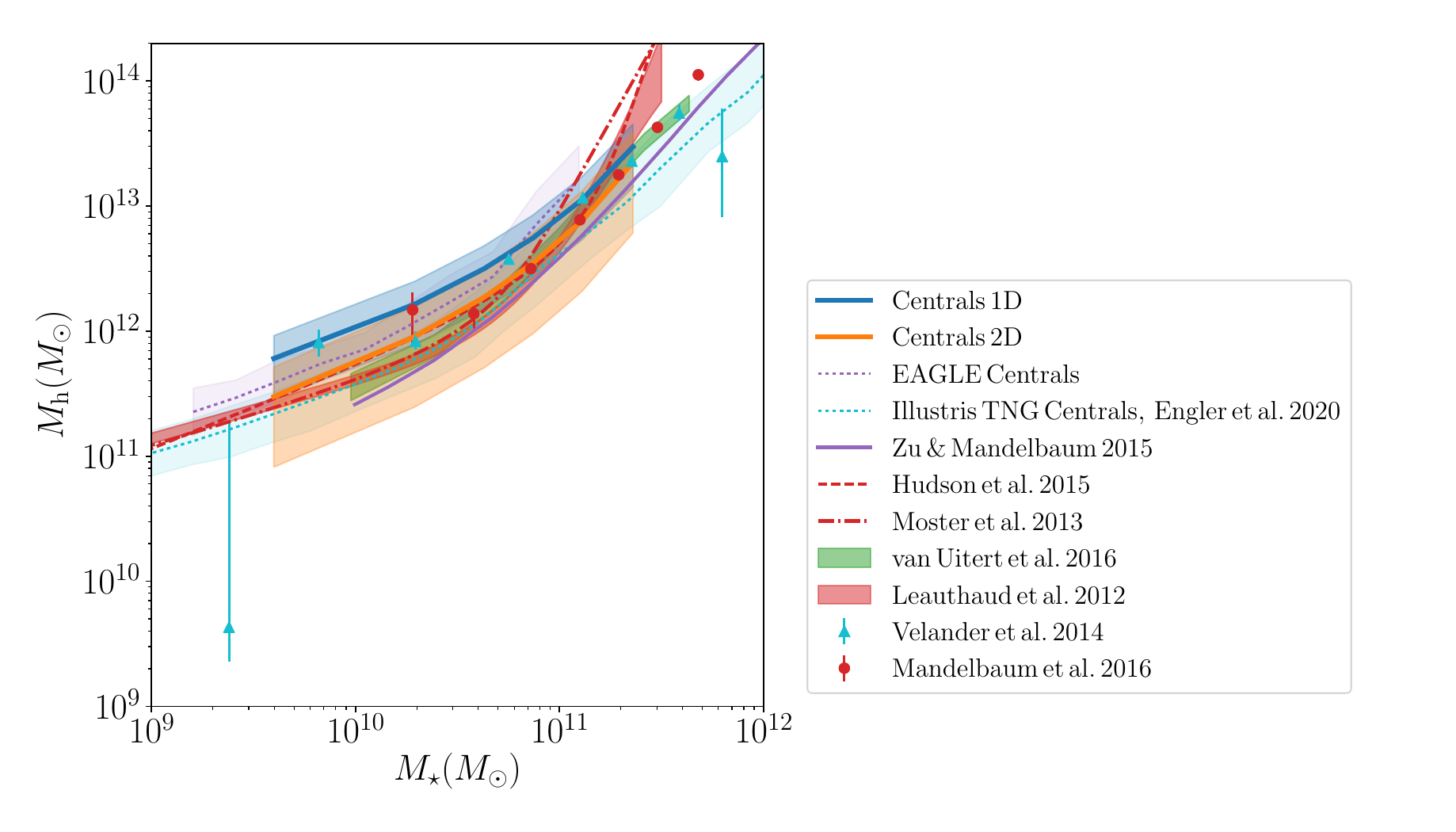}
        \caption{Stellar-to-halo mass relation for the central galaxies compared to the various published determinations of the relationship between the total and stellar mass of central galaxies. All of the relations between the total and stellar mass of central galaxies are in broad agreement and are all using galaxy-galaxy lensing to constrain the SMHM relation, excluding the simulations where those quantities are measured directly from the particle data.}
        \label{fig:smhm_cen_comp}
\end{figure*}

In Fig. \ref{fig:smhm_cen_comp} we show various published determinations of the relationship between the total and stellar mass of central galaxies \citep{Leauthaud2012, Moster2013, Velander2014, Hudson2015, Zu2015a, vanUitert2016, Mandelbaum2015a} as well as the EAGLE and Illustris TNG simulations \citep{Engler2020}. We scale all of these relations to our adopted values of $H_0$ and the definition of halo mass, that is to say the halo mass is defined with respect to 200 times the average density in the Universe, as is adopted in this paper. Furthermore, we also compare our results with the central and satellite properties from the hydrodynamical simulation EAGLE \citep{Schaye2015, McAlpine2015}. Specifically, we use the AGN model, from which we select galaxies with stellar masses ranging from $10^{9.6} M_{\odot}$ to $10^{11.2} M_{\odot}$ and their halo properties from which we can plot the mean SMHM relation and its scatter.

All of the relations between the total and stellar mass of central galaxies are in broad agreement and they all use galaxy-galaxy lensing to constrain the SMHM relation \citep[in the case of][also in combination with galaxy clustering and the stellar mass function, respectively]{Zu2015a, vanUitert2016}. For \citet{Velander2014} and \citet{Mandelbaum2015a}, we show their SMHM relation of red galaxies; as in the GAMA sample, central galaxies are mostly identified as red. Our measurements are also in agreement with the previous results. We also need to point out, as mentioned by \citet{Sifon2018}, that the measurements from \citet{Zu2015a} and \citet{Mandelbaum2015a} agree perfectly when the galaxies between the two samples are matched. The SMHM relations of \citet{Leauthaud2012}, \citet{Moster2013}, and \citet{Hudson2015} can be considered as better comparisons than the ones from the EAGLE or Illustris TNG simulations since they are obtained at redshifts comparable to the redshifts of the GAMA galaxies.

\begin{figure*}
        \includegraphics[width=\textwidth]{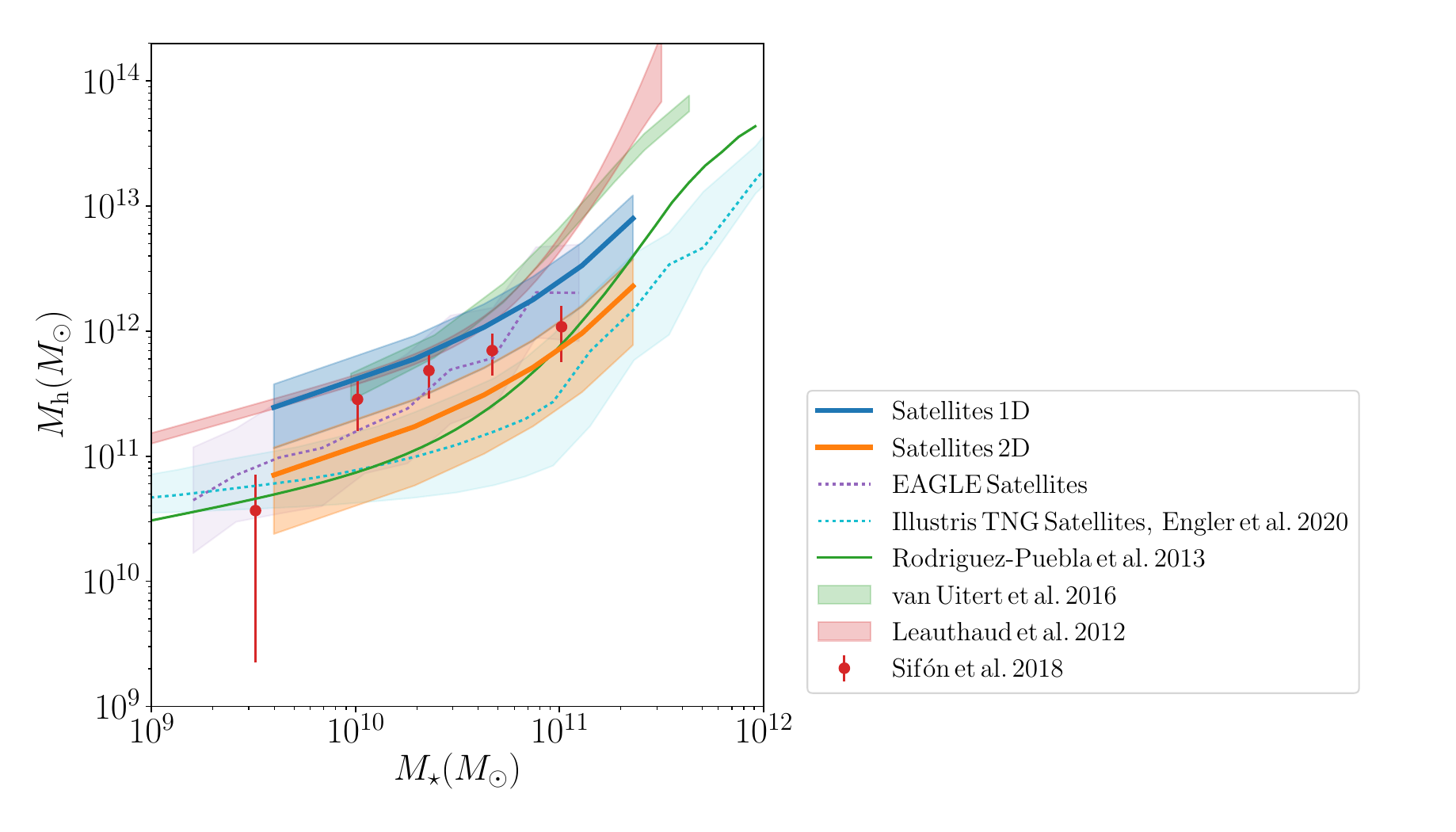}
        \caption{Stellar-to-halo mass relation for the satellite galaxies compared to the various published determinations of the relationship between the total and stellar mass of satellite galaxies. We show the SMHM relations of \citet{Leauthaud2012} and \citet{vanUitert2016} for comparison purposes with the centrals since they capture the majority of the other relations shown in Fig. \ref{fig:smhm_cen_comp}.}
        \label{fig:smhm_sat_comp}
\end{figure*}

\begin{figure}
        \includegraphics[width=\columnwidth]{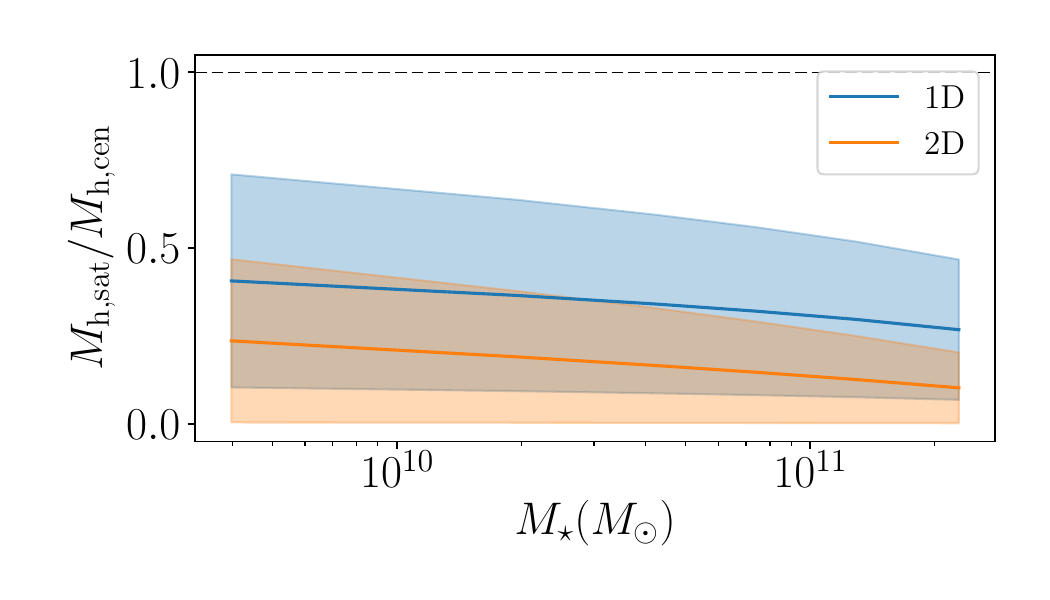}
        \caption{Ratio between the satellite and central halo mass as a function of stellar mass.}
        \label{fig:ratio}
\end{figure}

Similarly, for satellites, we show in Fig. \ref{fig:smhm_sat_comp} the comparison of the results from \citet{Rodriguez-Puebla2013}, \citet{Sifon2018}, and the EAGLE simulation with our two methods. However, the definitions of the halo mass of satellite galaxies in both \citet{Rodriguez-Puebla2013} and \citet{Sifon2018} are not equivalent to the one we use throughout this paper and it is also hard to correct this in order to compare the same quantities. \citet{Rodriguez-Puebla2013} use the definition of the subhalo mass that is defined as a mass of a satellite halo at the observed time (present time), and \citet{Sifon2018} use the definition of the subhalo mass as the mass within  a radius for which the subhalo density matches the background density of the cluster at the distance of the subhalo in question. The closest definition to ours is the definition from the EAGLE simulation. Our results are similar to the behaviour of satellite galaxies therein. As for the comparison with \citet{Rodriguez-Puebla2013} and \citet{Sifon2018}, all of the studies show lower satellite masses compared to the central galaxies at the same stellar mass. The same also holds true for the overall trend as a function of stellar mass. In Fig. \ref{fig:ratio} we also show the ratio between the satellite and central halo mass as a function of stellar mass. We observe that for the low stellar mass galaxies, the ratio is around 0.4 and drops towards 0.2 for high stellar mass galaxies, although the uncertainties on the ratio are quite large. This result directly shows us that more than \textasciitilde80\% of the dark matter of satellite galaxies is stripped (but with a large uncertainty), when they are accreted by a massive central galaxy.

Some studies of cosmological N-body simulations find that the ratio of satellite galaxy masses over central galaxy masses is $M_{\mathrm{h,sat}} / M_{\mathrm{h,cen}} \ll 1$ \citep[][and the references therein]{VandenBosch2018}, which means that the majority of dark matter haloes that merge with a central halo would be completely tidally disrupted. The tidal stripping and disruption of subhaloes in numerical simulations was mostly shown to be of numerical origin, emphasising that the cosmological simulations still suffer from overmerging \citep{VandenBosch2018}. Even though our ratio of halo masses between the satellites and centrals show large uncertainties, with a refined analysis, it would be possible to use this type of measurement as an independent observational confirmation for the lack (or presence) of numerical convergence in the simulations. Given our results, as presented in Fig. \ref{fig:ratio}, we cannot make a clear statement, although the results hint that less stripping is present in the observational data. On the other hand, the hydrodynamical simulations, such as EAGLE, show that the artificial stripping of haloes is not significant \citep{Chaves-Montero2016}, but we need to caution for the survivor bias in this case since completely disrupted haloes would not be present in the sample.

All of the previous results, together with our findings show that the satellite galaxies are indeed preferentially stripped of their dark matter and the effect can be observed as higher stellar mass to halo mass ratios compared to their central counterparts of a similar stellar mass. All of the measurements show a statistically different SMHM relation for the central and satellite galaxies, which furthermore shows that two-dimensional galaxy-galaxy lensing can measure the SMHM relation for different populations of galaxies.

\section{Discussion and conclusions}
\label{sec:conclusions}

We have measured the stellar-to-halo mass relation of central and satellite galaxies in the GAMA survey. In this analysis, we use the more advanced two-dimensional galaxy-galaxy lensing method to constrain the SMHM relation, which has potential benefits over the traditionally used stacked tangential shear method \citep[also referred to as one-dimensional galaxy-galaxy lensing here,][]{Dvornik2019}.

We use the three equatorial GAMA patches that overlap with the KiDS data in order to measure both the tangential shear signal around the central and satellite galaxies as well as the two-dimensional galaxy-galaxy lensing constraints on the same lenses and sources. The shear signals are then used to constrain the SMHM relation of central and satellite galaxies. 

We model the lensing signal using an NFW profile together with the concentration-mass relation by \citet{Duffy2011}, scaled by a normalisation factor for which we took the fit into account. We assume a functional form for the SMHM relation in the form of an exponential function, which was motivated by the observed behaviour in the simulations, and we fold it through our model, thus directly fitting for the normalisation and slope of the SMHM relation. The lens model is used to calculate the tangential shear profile, which is then fitted to the measured tangential shear profile from the GAMA and KiDS data, as well as to directly predict the two Cartesian components of the galaxies' ellipticities used in our two-dimensional method. 

We find that the SMHM relation can be successfully measured using the two-dimensional method, with a comparable statistical power to the traditional one-dimensional method using the stacked tangential shear measurements. Both methods give us similar results for the SMHM relations, showing that the two-dimensional method is indeed a robust way to measure properties of the galaxy--halo connection, without using statistically equivalent samples as in the case of the one-dimensional method, nor using more complicated halo models or relying on support from other probes. The resulting SMHM relations are broadly in agreement with the literature, and our results show that the satellite galaxies are indeed preferentially stripped of their dark matter and the effect can be observed as higher stellar mass to halo mass ratios compared to their central counterparts of a similar stellar mass.

By comparing the results of this paper with the findings of our previous paper \citep{Dvornik2019}, we are able to recognize that the comparable constraining power of the one-dimensional and two-dimensional method, shown in Fig.\ref{fig:corner}, is unexpected. In the \citet{Dvornik2019} paper, we predicted a factor of 3 improvement. As seen in the results, the statistical powers of both methods are comparable. \citet{Dvornik2019} explores an  idealised mock dataset as well as noiseless and complete simulations, where the exact galaxy classification was also known. As mentioned in Sec. \ref{sec:results}, multiple effects can and will cause differences from this idealised mock. First, the increased uncertainty of the two-dimensional method is directly dependent on how well one can identify central and satellite galaxies and how robust this identification and selection is. Even though the GAMA group catalogue \citep{Robotham2011} is highly robust, this does not mean it is perfect and even a small number of incorrect classifications of galaxies would cause excessive scatter in the resulting SMHM relations, impacting the ability of the two-dimensional model to constrain the parameters. Secondly, the results of the two-dimensional galaxy-galaxy lensing seems to be biased due to the completeness limit of the GAMA survey. We have shown that slight incompleteness of the lens sample can cause biases in the inferred parameters, which can be as large as 10\% (as is shown in Fig. \ref{fig:mice}). This can be somewhat seen in Fig. \ref{fig:corner}, where the $\gamma_{\mathrm{cen}}$ parameters are most noticeably different. The two-dimensional analysis is still computationally and resource expensive compared to the one-dimensional method. This further reduces the usability of the method, and our preferred galaxy-galaxy lensing analysis thus remains the standard one-dimensional approach.

\begin{acknowledgements}
We thank the anonymous referee for their very useful comments and suggestions. AD, AHW and HHi acknowledge support from ERC Consolidator Grant (No. 770935). HHi is further supported by a Heisenberg grant of the Deutsche Forschungsgemeinschaft (Hi 1495/5-1). HHo and AK acknowledge support from Vici grant 639.043.512, financed by the Netherlands Organisation for Scientific Research (NWO). KK acknowledges support by the Alexander von Humboldt Foundation. MB is supported by the Polish Ministry of Science and Higher Education through grant DIR/WK/2018/12, and by the Polish National Science Center through grant no. 2018/30/E/ST9/00698. CH, MA, CL, BG and TT acknowledge support from the European Research Council under grant number 647112, and CH further acknowledges support from the Max Planck Society and the Alexander von Humboldt Foundation in the framework of the Max Planck-Humboldt Research Award endowed by the Federal Ministry of Education and Research. TT acknowledges funding from the European Union’s Horizon 2020 research and innovation programme under the Marie Skłodowska-Curie grant agreement No 797794.\\

This research is based on data products from observations made with ESO Telescopes at the La Silla Paranal 
Observatory under programme IDs 177.A-3016, 177.A-3017 and 177.A-3018, and on data products produced 
by Target/OmegaCEN, INAF-OACN, INAF-OAPD and the KiDS production team, on behalf of the KiDS consortium.\\

GAMA is a joint European-Australasian project based around a spectroscopic campaign using the Anglo-Australian Telescope. 
The GAMA input catalogue is based on data taken from the Sloan Digital Sky Survey and the UKIRT Infrared Deep Sky Survey. 
Complementary imaging of the GAMA regions is being obtained by a number of independent survey programs including GALEX MIS, 
VST KiDS, VISTA VIKING, WISE, Herschel-ATLAS, GMRT and ASKAP providing UV to radio coverage. GAMA is funded by the 
STFC (UK), the ARC (Australia), the AAO, and the participating institutions. The GAMA website is \mbox{\url{http://www.gama-survey.org}}.\\

This work has made use of Python (\url{http://www.python.org}), including the packages \texttt{numpy} (\url{http://www.numpy.org}) 
and \texttt{scipy} (\url{http://www.scipy.org}). Plots have been produced with \texttt{matplotlib} \citep{Hunter2007}. This work has made use of CosmoHub. CosmoHub has been developed by the Port d'Informació Científica (PIC), maintained through a collaboration of the Institut de Física d'Altes Energies (IFAE) and the Centro de Investigaciones Energéticas, Medioambientales y Tecnológicas (CIEMAT) and the Institute of Space Sciences (CSIC \& IEEC), and was partially funded by the “Plan Estatal de Investigación Científica y Técnica y de Innovación” program of the Spanish government. \\
\\
\textit{Author contributions:} All authors contributed to writing and development of this paper. The authorship list reflects the lead authors (AD, HHo, KK) followed by two alphabetical groups. The first alphabetical group includes those who are key contributors to both the scientific analysis and the data products. The second group covers those who have made a significant contribution either to the data products or to the scientific analysis.
\end{acknowledgements}

\bibliographystyle{aa}
\bibliography{library} 

\end{document}